\documentclass[12 pt]{article}
\usepackage{amsmath,amssymb,braket}
\usepackage[english]{babel}
\usepackage{lipsum}
\usepackage{hyperref}
\date{}
\begin{document}
\title{\textbf{A SHORT SURVEY AROUND THE PUMPING LEMMA FOR CONTEXT-FREE LANGUAGES}}
\author{\sc Gabriele Gull\`a}
\maketitle
\begin{abstract}Following a seminar the present author gave to an Automata Theory course to computer science students, it will be presented - in a very synthetic and mostly self-contained way - the principal properties of \textit{context-free languages} (CFL) - with particular attention given to the \textit{Pumping Lemma} (PL) - and of grammars which generate them (CFG). We refer to Chomsky and Sch\"utzenberger for the first works about it.\\
What is known in literature as the \textit{Iteration Theorem} here will be referred to as the \textit{Ogden's Lemma} in a fully justified way.\\
All definitions not strictly connected with the notion of ``context-freeness'' will be omitted (we will give precise references for all of them).\\
The symbology used is substantially the classical one, but we will replace some symbols to avoid confusion with those used in logic: in particular it will be used for \textit{derivation} and for its transitive and reflexive closure the symbols ``$\rightharpoonup$'' and ``$\stackrel{*}\rightharpoonup$''. 
\end{abstract}
\let\thefootnote\relax
\section{Context-free languages and grammars}
We start with the basic notions which will recur in the rest of the paper, in particular CFL and CFG, and some examples.
\\\\\textbf{Definition 1.1}\\
Let $\mathcal{G}$ be a grammar; we set as:\\
1) $V$ the set of its \textit{non terminal symbols} (or \textit{variables});\\
2) $T$ the set of \textit{terminal symbols};\\
3) $P$ the finite set of \textit{productions};\\
4) $S\in V$ the \textit{axiom} of the grammar;\\
5) $\Sigma: =V\cup T$;\\
Then $\mathcal{G}$ is a CFG if and only if it holds
$$p\in P\Leftrightarrow p=A\rightarrow \alpha$$
where $A\in V$ and $\alpha\in \Sigma^*$ (the \textit{transitive and reflexive closure} of $\Sigma$).\\ See \cite{HU} for accessory definitions.\\
\\As known by \textit{Chomsky hierarchy}, CFG $\underset{\neq}\supset\left\{\mbox{\textit{Regular} Grammars}\right\}$.\\
\\\textbf{Remark 1.1}\\
The locution ``context-free'' comes from the fact that the first member of productions of a CFG is never juxtaposed to some $\beta\in \Sigma^*$.\\
About this let us consider the following \textbf{Example 1.1}:\\
let be $\mathcal{L}=\left\{a^nb^nc^n| n\ge 1\right\}$; this language can be generated by the grammar such that\\
1) $T=\left\{a, b, c\right\}$\\
2) $V=\left\{S, B\right\}$\\
3) $P=\left\{S\rightarrow aBSc, S\rightarrow abc, Ba\rightarrow aB, Bb\rightarrow bb\right\}$\\
so, in the last production, $B$ can be replaced by $b$ only if on its right side it appears the ``context'' $b$. Then $\mathcal{L}$ is a \textit{context-sensitive} language. We will see in what follows that $\mathcal{L}$ is strictly context-sensitive, meaning that\\ CFG $\underset{\neq}\subset \left\{\mbox{Context-sensitive Grammars}\right\}$.\\
\\Obviously $\mathcal{L}$ is a CFL if and only if it is generated by a CFG, meaning its elements are words of $T^*$ \textit{derived} ($\stackrel{*}\rightharpoonup$) from $S$ in some CFG.\\
An element $\beta\in \Sigma^*$ is said to be a \textit{sentential form} if $S\stackrel{*}\rightharpoonup \beta$.
\\\\\textbf{Example 1.2}\\
Let $\mathcal{G}$ be a grammar such that:\\
1) $V=\left\{S\right\}$;\\
2) $T=\left\{a, b\right\}$;\\
3) $P=\left\{(I) S\rightarrow aSb, (II) S\rightarrow ab\right\}$\\
As the left member of every production contains just a variable, this is a CFG.\\ 
The generated language contains surely the word $a^nb^n\quad \forall\enspace n\ge 1$, because by applying $n-1$ times the production (I) and then (II) we obtain
$$S\rightarrow aSb\rightarrow\cdot \cdot \cdot \cdot \cdot\rightarrow a^{n-1}Sb^{n-1}\rightarrow a^nb^n$$
Moreover it does not contain words of other form, in fact only production (II) can be used alone (and it produces $ab$), while production (I) can be used an arbitrary number of times (generating strings containing always the symbol $S$), but this derivation has to be always completed by a unique use of (II).\\
So the CFL generated is $\mathcal{L}=\left\{a^nb^n| n\ge 1\right\}$. 
\begin{flushright}$\mathcal{a}$\end{flushright}
\section{Derivation trees}
Now let us see another representation of derivation relation. We have to introduce some ideas from graph theory. The neophyte interested reader has an almost infinite choice about the literature on graphs: my suggestion is ``Modern graph theory'', by B\'ela Bollob\'as. \\
\\\textbf{Definition 2.1}\\
A (finite) \textit{tree} is a \textit{minimally connected} graph, meaning that if one eliminates an edge between two nodes, one obtains two indipendent graphs.\\
If we set a node as the initial one we obtain a \textit{rooted} tree. In this case the edges assume an ideal orientation from the root-node.\\The nodes of a rooted tree can be ordered (and so enumerated) firtsly starting from the root to the low, and then from left to right. Given a node $n$, the nodes directly connected to this and labeled by a bigger ordinal, are called \textit{sons} of the node.\\
Every node in a tree is characterized by another natural number, the \textit{degree} of the node, which indicates the number of outgoing and ingoing edges (from and to the node).\\
So we say that a node is \textit{internal} if its degree is $\ge 2$, while if the degree is 1 the node is called a \textit{leaf} (or \textit{external node} or \textit{terminal}).\\In a tree a \textit{path} is a concatenation of edges and its  \textit{length} is the number of this edges.\\
\\\textbf{Definition 2.2}\\
Given a CFG, a rooted tree is a \textit{derivation} one if:\\
1) every node is labeled by a symbol in $\Sigma\cup \left\{\epsilon\right\}$;\\
2) the label of the root is the axiom $S$ of the CFG;\\
3) internal nodes are labeled by elements of $V$;\\
4) if the node $n$ is labeled by $A$ and its sons are $n_1, n_2, ...., n_k$ with labels $X_1, X_2,....., X_k$, then $A\rightarrow X_1X_2.....X_k$ is a production of the grammar.\\
5) If a node $n$ has as a label the empty symbol $\epsilon$, then this node is a leaf and it is the only son of $n$. \\
\\Let $T$ be a rooted tree; then $T'$ is a \textit{subtree} of $T$ if $T'$ is a rooted tree, $T'\subset T$ but its root can be different from $S$.\\
Finally we call \textit{yield} of a tree $T$ the word obtained reading labels of the leaf of $T$ following the order settled above: the higher leaves have the precedence, and between two leaves of the same height, the one on the left has the precedence.\\
\\All this in order to obtain the following\\
\\\textbf{Theorem 2.1}\\
Given a CFG, $\beta$ is a sentential form if and only if there is a derivation tree with yield $\beta$.
\begin{flushright}$\mathcal{a}$\end{flushright}
I omit the proof (see for example \cite{HU}, cap. 4 sec. 3), but I want to stress the strong relationship between derivations and derivation trees which will allow to prove the PL.\\
\\\textbf{Remark 2.1}\\
By definition, a derivation tree does not determine univocally the order that must be followed by the application of the productions.
\section{CFG normal form}
Here we are interested in a particular procedure which transforms any CFG in an equivalent one from the generative power perspective, but whose productions assume just two precise forms.\\
In this section I will not present complete proofs but just some hints. For further details see \cite{H} and \cite{HU}.\\
\\\textbf{Definition 3.1}\\
We say that a symbol $X\in \Sigma$ is \textit{usefull} if there is a derivation\\ $S\stackrel{*}\rightharpoonup \alpha X\beta\stackrel{*}\rightharpoonup u$, with $\alpha, \beta\in \Sigma^*$ and $u\in T^*$.\\If there is not such a derivation the symbol is said to be \textit{useless}.\\
A production of the type $A\rightarrow \epsilon$ is called \textit{$\epsilon$-production}.
\\Productions of type $A\rightarrow B, B\in V$ are called \textit{unitary productions}.
\\\\Then:\\
\\\hypertarget{*}{\textbf{Lemma 3.1}}\\
Every non empty CFL without $\epsilon$ is generated by a CFG without useless symbols, $\epsilon$-productions or unitary productions.\\
\\\textbf{Proof} (idea)\\
This result needs several other lemmas in order to be proven, so I will give just some hints about the removal of useless symbols. For the $\epsilon$-productions and unitary productions the reader can check any textbook about CFG.\\
Let $\mathcal{G}=(V, T, P, S)$ be a CFG; then there is an effective procedure (based on an iterative algorithm and completed with an induction) which allows to construct a CFG $\enspace\mathcal{G}'=(V', T, P', S)$ such that for every $A\in V'$ there is some $u\in T^*$ such that $A\stackrel{*}\rightharpoonup u$.\\
Now, thanks to $\mathcal{G}'$ we can always obtain (in an effective way) a CFG \\$\mathcal{G}''=(V'', \tilde{T}, P'', S)$ such that for every $X\in (\Sigma'')$ there are $\alpha, \beta \in (\Sigma'')^*$ such that $S\stackrel{*}\rightharpoonup \alpha X\beta$.
\begin{flushright}$\mathcal{a}$\end{flushright}
From the previous lemma we obtain the following \\
\\\hypertarget{**}{\textbf{Theorem 3.1}}\\
Every CGL without $\epsilon$ is generated by a CFG whose productions are of the following kind\\
(I) $A\rightarrow BC$ or\\
(II) $A\rightarrow a$\\
with $B, C\in V$ and $a\in T$.\\
This particular form of CFG is called \textit{Chomsky normal form} (CNF).
\begin{flushright}$\mathcal{a}$\end{flushright}
\textbf{Remark 3.1}\\
In order to obtain a CFG without useless symbols, the algorithms $\mathcal{G}\longmapsto \mathcal{G'}$ and $\mathcal{G'}\longmapsto \mathcal{G''}$ used in proof of \hyperlink{*}{Lemma 3.1} have to run exactly in that order.\\
\\Let us see why with the following\\
\\\textbf{Example 3.1}\\
Let us consider a CFG with productions\\
1) $S\rightarrow AB$\\
2) $S\rightarrow a$\\
3) $A\rightarrow a$\\
and we cancel the useless symbols:\\
step (I)- because from $B$ one can not derive any terminal symbol, we can eliminate it together with the productions which involve it. So we reduce to \\
1') $S\rightarrow a$\\
2') $A\rightarrow a$\\
step (II)- now the unique production such that there is a derivation $S\stackrel{*}\rightharpoonup \gamma$ is the (1'), so we can eliminate (2') and with that the useless symbol $A$ obtaining an CFG equivalent to the given one.\\
By applying step (II) as the first one we would conclude that just $B$ is useless, because $A$ apperas in the production of the sentential form $a$:\\ ($S\rightarrow AB\rightarrow a$).\\
By applying now step (I) we should keep the production $A\rightarrow a$ and with this the useless symbol $A$.
\begin{flushright}$\mathcal{a}$\end{flushright}
The following is a simple example about passing from a CFG to another in CNF:
\newpage\noindent\textbf{Example 3.2}\\
Let us consider a CFG defined by the productions\\
1) $S\rightarrow bA$\\
2) $S\rightarrow aB$\\
3) $A\rightarrow bAA$\\
4) $A\rightarrow aS$\\
5) $A\rightarrow a$ \\
6) $B\rightarrow aBB$ \\
7) $B\rightarrow bS$ \\
8) $B\rightarrow b$\\
we do not touch productions (5) and (8) (they are already in the correct form).\\
In the other productions we replace, on the right, the terminal symbols with variables:\\
1) $S\rightarrow bA$ becomes $S\rightarrow CA$ and $C\rightarrow b$\\
2) $S\rightarrow aB$ becomes $S\rightarrow DB$ and $D\rightarrow a$\\
3) $A\rightarrow bAA$ becomes $A\rightarrow CAA$\\
4) $A\rightarrow aS$ becomes $A\rightarrow DS$\\ 
6) $B\rightarrow aBB$ becomes $B\rightarrow DBB$\\
7) $B\rightarrow bS$ becomes $B\rightarrow CS$\\
now we transform productions which have more than two variables on the right:\\
$A\rightarrow CAA$ becomes $A\rightarrow CE$ and $E\rightarrow AA$\\
$B\rightarrow DBB$ becomes $B\rightarrow DH$ and $H\rightarrow BB$\\
So the new CFG (again: equivalent to the given one) is defined by the new productions:\\
1) $S\rightarrow CA$\\
2) $C\rightarrow b$\\
3) $S\rightarrow DB$\\
4) $D\rightarrow a$\\
5) $A\rightarrow DS$\\
6) $B\rightarrow CS$\\
7) $A\rightarrow CE$\\
8) $E\rightarrow AA$\\
9) $B\rightarrow DH$\\
10) $H\rightarrow BB$\\
11) $A\rightarrow a$\\ 
\\12) $B\rightarrow b$
\begin{flushright}$\mathcal{a}$\end{flushright}
\textbf{Remark 3.2}\\
The CNF is not the only normal form for CFG. Another one, extremely important (but which I will not present here), is the \textit{Greibach normal form}, for which there is a theorem similar to \hyperlink{**}{Theorem 3.1}.\\
One of the interesting things about this form is the possibility to use it in order to obtain set-theoretic and algebraic representations for CFG: see in \cite{ABeBo} theorems by Shamir, Chomsky-Sch\"utzenberger, Greibach and Wechler.
\section{The Pumping Lemma for CFL}
Now we are ready to treat the Pumping Lemma, whose first proof appeared in Bar Hillel-Perles-Shamir\cite{BHPS}. This result is an almost straightforward corollary of Ogden's Lemma (for this one see \cite{H} and \cite{HU}) proved in 1968; here I present a direct proof of PL.\\
We start from a lemma which will help us to prove PL.\\
\\\hypertarget{***}{\textbf{Lemma 4.1}}\\
Let us consider a CFG in CNF and the associated derivation tree, let $u$ be the yield of this tree and let $n$ be the length of the longer path in the tree. Then $|u|\le 2^{n-1}$.\\
\\\textbf{Proof}\\
By induction on $n$:\\
1) if $n=1$ then the tree is composed by two nodes, the root $S$ and the terminal $u$, connected by an edge. Then $|u|=1=2^{1-1}$ and the based of induction is proved.\\
2) If $n>1$ then the tree starts with a production of the type $A\rightarrow BC$ and not with a production which produces a terminal symbol. Let us consider the rooted subtrees in $B$ and $C$: in these the longest path will be surely shorter than $n$, so by the inductive hypothesis their havests have length not bigger than $2^{n-2}$.\\
The yield of the entire tree is the concatenation of this last two yields, so its length is not bigger than $2^{n-2}+2^{n-2}=2^{n-1}$.
\begin{flushright}$\mathcal{a}$\end{flushright}
\textbf{Pumping Lemma for CFL (Bar Hillel-Perles-Shamir, 1961)}\\
Let $\mathcal{L}$ be a CGL. Then there is a constant $N_{\mathcal{L}}$ such that if $z\in \mathcal{L}$ and $|z|\ge N_{\mathcal{L}}$, then one has that $z=uvwxy$ such that\\
1) $|uwx|\le N_{\mathcal{L}}$\\
2) $|vx|\ge 1$\\
3) $\forall \enspace i\ge 0\quad uv^iwx^iy\in \mathcal{L}$\\
\\\textbf{Proof}\\
Let $\mathcal{L}$ be a CFL; we can suppose $\left\{ \epsilon\right\}\neq \mathcal{L}\neq \emptyset$ otherwise the thesis is trivially verified.\\
Then let $\mathcal{G}$ be a CFG in CNF which generates $\mathcal{L}\setminus \left\{ \epsilon\right\}$, and let $m$ be the number of variables of $\mathcal{G}$.\\
We chose $N_{\mathcal{L}}=2^m$. We know from \hyperlink{***}{Lemma 4.1} that a tree whose longest path has length $m$ must have a yield not bigger than $2^{m-1}=\frac{N_{\mathcal{L}}}{2}$. Now, $z$ is too long to be the yield of such a tree, so the tree associated to the yield $z$ (we call it $T$) must have the longest path of length at least $m+1$.\\
Such a path will contain at least $m+1$ variables $A_0, A_1,....., A_m$, and amongst them there will be two equal; we call them $A_i=A_j$ with $i< j$. Then the subtree $T_j$ rooted in $A_j$ will be strictly contained in the subtree $T_i$ rooted in $A_i$ (because the latter precedes $A_j$) and we can take as $w$ the yield of $T_j$, as $v$ and $x$ the strings respectively on the left and on the right of $w$ in the tree $T_i$, and as $u$ and $y$ the strings respectively on the left and on the right of $v$ and $x$ in $T$. We notice that, since in a CFG in CNF there are not unitary productions, $v$ and $x$ can not be both $\epsilon$ and so $|vx|\ge 1$. So we have shown (2) and (3), but this last just for $i=1$. For $i=0$ it is enough to eliminate $T_i$ and pose $A_i$ as the root of $T_j$: we can do it because the two roots are labeled by the same symbol. Then in this case $v$ and $x$ disappear and we obtain that $uv^0wx^0y\in \mathcal{L}$.\\
For $i>1$ it is enough to eliminate $T_j$ and pose $A_j$ as the root of $T_i$ (legitimate thing, as already seen): in this way we obtain $uv^2wx^2y$; by iterating this replacement of trees, one obtains all the other $uv^iwx^iy$.\\
There is still to prove (1):\\
it is obvious that, as $T_i\subseteq T$, its longest path is no greater than $m+1$, so, again from \hyperlink{***}{Lemma 4.1}, its yield is $vwx$ and it will have length no greater than $2^m=N_{\mathcal{L}}$.
\begin{flushright}$\mathcal{a}$\end{flushright}
As in the regular case, also in the context-free case the PL can be used to establish if a language is not a CFL. What follows is an example about this:
\\\\\textbf{Example 4.1}\\
We prove that $\mathcal{L}=\left\{a^{2^n}| n\ge 1\right\}$ is not a CFL.\\
Let us suppose \textit{ad absurdum} that it is, and let $N$ be the constant of PL.\\We notice that every word in $\mathcal{L}$ has as length a power of 2. So let $z=a^{2^N}=uvwxy$ be a word such that surely $|z|=2^N> N$ and with $|vx|\ge 1$ and $|vwx|\le N$.\\
Then
$$|vx|=2^aB,\enspace 0\le a\le (N-1) \wedge B\in\left\{2m+1|\forall \enspace m\in \mathbb{N}\right\}$$
meaning it is a power of 2, or it is an even number or an odd one.\\
But then, in every case
$$|uwy|=2^N-2^aB=2^a(2^{N-a}-B)\neq 2^k\quad \forall \enspace k\in \mathbb{N}$$
\\So $uwy\notin \mathcal{L}$: contradiction.
\begin{flushright}$\mathcal{a}$\end{flushright}

The classic example, presented in almost every textbook, about the use of PL (see for example \cite{HU}) is
$$\mathcal{L}=\left\{a^nb^nc^n|n\ge 1\right\}$$
that we already saw in the first section.\\ 
If $v$ and $x$ contain at least two different symbols, then $uv^2wx^2y$ contain mixed symbols, so it is not in $\mathcal{L}$: contradiction.\\
If conversely $v$ and $x$ contain just one symbol, then $uv^2wx^2y$ does not contain the same number of $a, b$ and $c$, and again it is not  in $\mathcal{L}$. Again a contradiction.\\
\begin{flushright}$\mathcal{a}$\end{flushright}
I will use this language in the next section.
\section{Some properties of closure and decidability}
In this section I will present some particular properties of CFL still connected to the PL.\\
For the nature of PL itself, there are cases such that this result is not helpfull, for instance when the language satisfies the thesis of PL even if it is not a CFL. Then it is possible to show that a given language is not a CFL by using the Ogden's Lemma, or generating from this language another non context-free one by using just operations which preserve CFL.\\ 
\\\textbf{Definition 5.1}\\
Let $A$ and $B$ be two alphabets. A \textit{substitution} is a function $f:A\longrightarrow \mathcal{P}(B^*)$ which associates to every symbol of $A$ a language in $B^*$. This function extends to words by posing:\\
$*$) $f(\epsilon)=\epsilon$;\\
$*$) $f(ua)=f(u)f(a)$\\
and then to languages by posing $f(\mathcal{L})=\bigcup_{x\in \mathcal{L}}f(x)$.\\
\\\textbf{Theorem 5.1}\\
The set of CFL is closed under the following operations:\\
1) Union\\
2) Concatenation\\
3) Kleene closure ($\mathcal{L}^*=\bigcup_{i=0}^{\infty} \mathcal{L}^i$)\\
4) Substitution\\
\\\textbf{Proof} (idea)\\
Let $\mathcal{L}_1$ and $\mathcal{L}_2$ be two CFL generated by $\mathcal{G}_1=(V_1, T_1, P_1, S_1)$ and $\mathcal{G}_2=(V_2, T_2, P_2, S_2)$. Then (1) the grammar
$$\mathcal{G}_3=(V_1\cup V_2\cup \left\{S_3\right\}, T_1\cup T_2, P_1\cup P_2\cup \left\{S_3\rightarrow S_1|S_2\right\}, S_3)$$
generates $\mathcal{L}_3=\mathcal{L}_1\cup \mathcal{L}_2$.\\
(2) the grammar
$$\mathcal{G}_4=(V_1\cup V_2\cup \left\{S_4\right\}, T_1\cup T_2, P_1\cup P_2\cup \left\{S_4\rightarrow S_1S_2\right\}, S_4)$$
generates $\mathcal{L}_4=\mathcal{L}_1\mathcal{L}_2$.\\
(3) the grammar
$$\mathcal{G}_5=(V_1\cup \left\{S_5\right\}, T_1, P_1\cup \left\{S_5\rightarrow S_1S_5|\epsilon \right\}, S_5)$$
generates $\mathcal{L}_5=\mathcal{L}_1^*$.\\
By definition of substitution, (4):\\
let $\mathcal{G}$ and $\mathcal{G}_a$ be the CFG which generate the CFL $\mathcal{L}$ and $\mathcal{L}_a$, with $a\in \Sigma$. The language $f(\mathcal{L})$ is generated by the grammar $\tilde{\mathcal{G}}$ constructed in this way:\\
\\$\bullet\qquad \tilde{V}=V\cup\left\{\mathcal{G}_a\right\}_{a\in \Sigma}$ \\
$\bullet\qquad \tilde{T}=\left\{T_a\right\}_{a\in \Sigma}$\\
$\bullet\qquad \tilde{S}= S$\\
$\bullet\qquad \tilde{P}=\left\{P_a\right\}_{a\in \Sigma}\cup \left\{P'\right\}$\\
\\where $P'$ contains productions of $\mathcal{G}$ of the form $A\rightarrow \alpha$, where every instance of $a\in \alpha$ are substituted by the axiom $S_a$.\\
\\Since a homomorphism is a particular substitution where $f(a)$ contains a single word, it follows also that CFL are closed under homomorphisms too.\\
It is possible to show that they are closed under inverse homomorphisms too, but the proof of this result uses \textit{pushdown automata}, the context-free couterpart of finite state automata (those associated to regular grammars).
\begin{flushright}$\mathcal{a}$\end{flushright}
\textbf{Theorem 5.2}\\
The set of CFL is not closed under intersection and complementation.\\
\\\textbf{Proof}\\
Let $\mathcal{L}_1=\left\{a^ib^ic^k|i\ge 1, k\ge 1\right\}$ and $\mathcal{L}_2=\left\{a^ib^kc^k|i\ge 1, k\ge 1\right\}$; this languages are generated respectively by the grammars with productions:\\
\\(1) $S\rightarrow AB,\qquad A\rightarrow aAB| ab,\qquad B\rightarrow cB| c$\\
(2) $S\rightarrow CD,\qquad C\rightarrow ac| a,\qquad D\rightarrow bDc| bc$\\
\\so $\mathcal{L}_1$ and $\mathcal{L}_1$ are CFL, but $\mathcal{L}_1\cap \mathcal{L}_2=\left\{a^ib^ic^i|i\ge 1\right\}$, which, as we know from the previous section, is not a  CFL.\\
If they were closed under complementation they would be closed also for intersections thanks to De Morgan laws,\newpage
$$\mathcal{L}_1\cap \mathcal{L}_2=(\mathcal{L}_1^c\cup \mathcal{L}_2^c)^c$$
\begin{flushright}$\mathcal{a}$\end{flushright}
I want to end this brief survey with a classical decidability result for CFL, deriving from PL:\\
\\\textbf{Theorem 5.3}\\
Given a CFL it is decidable to establish if it is (1) empty, or (2) infinite, and so if it is (3) finite and if (4) a given word belongs to it or not.\\
\\\textbf{Proof}\\
Let $\mathcal{G}=(V, T, P, S)$ be a CFG in CNF, and let $\mathcal{L}$ be the CFL associated; then let $N=|V|$.\\
From PL if $z=uvwxy\in \mathcal{L}$, with $|z|> 2^N$, then $uwy\in \mathcal{L}$ e $|uwy|\le 2^N$.\\
So if $\mathcal{L}\neq \emptyset$ we can find some strings of length at most $2^N$.\\
Now, in a CFG in CNF the productions transform a variable in a concatenation of two variables,  or they substitute a variable with a terminal, so a string of length $k$ is generated by a derivation of length not bigger than $2k-1$.\\
This because, in the worst hypothesis, we deal with a situation of the type
$$S\rightarrow A_1A_2\rightarrow A_{11}A_{12}A_2\rightarrow A_{11}A_{12}A_{21}A_{22}\rightarrow ....\rightarrow a_1a_2....a_k$$
so, excepted for the first couple of variable generated by the axiom of the CFG, every variable is followed by a production that produces two others until for every variable it is generated a unique terminal: $2k-1$ productions.\\ 
So in oder to verify that in $\mathcal{L}$ there is a string of length no greater than $2^N$ it is enough to analyze the derivations of length no greater than $2^{N+1}-1$ which are at most $|P|\cdot2^{N+1}$. So we have shown (1) and (4).\\
If $2^N<z\le 2^{2N}$, always from PL, the language is infinite (because it contains infinite strings $uv^iwx^iy$), and so, as previously, it is enough to consider derivations of length no geater than $2^{2(N+1)}-1$, whose number is upper bounded by $|P|\cdot (2^{2N+1})$, and to verify that some of this produce a string of terminals. This shows (2) and (3) (because of course if $\mathcal{L}\neq \emptyset$ and it is not infinite then it is finite).
\begin{flushright}$\mathcal{a}$\end{flushright}


\begin{thebibliography}{abc}
\bibitem {ABeBo} Autebert, J-M. - Berstel, J - Boasson, L. : ``Context-Free Languages and Pushdown Automata'', cap. 3, vol. 1 del ``Handbook of Formal Languages'', Rozemberg-Salomaa Ed. per Springer, 1997;
\bibitem {BHPS} Bar Hillel, Y - Perles, M - Shamir, E : ``On Formal Properties of Simple Phrase Structure Grammars'', Zeitschrift f\"ur Phonetik, Sprachwissenschaft und Kommunikationsforschung, 1961;
\bibitem {BeBo} Berstel, J - Boasson, L. : ``Context-Free Languages'', cap. 2 del ``Handbook of Theoretical Computer Science'', J. van Leeuwen Ed., 1990;
\bibitem {C} Chomsky, N. : ``On Certain Formal Properties of Grammars'',\\ 
Information and Control 2, 1959;
\bibitem {CS} Chomsky, N. - Sch\"utzenberger,M.P. : ``The Algebraic Theory of Context-Free Languages'', in ``Computer Programming and Formal Languages'', Braffort-Hirschberg Ed. per  North Holland, 1963;
\bibitem {DLDA} De Luca, A - D'Alessandro, F. : ``Teoria degli Automi Finiti'',\\ 
Springer Ed., 2013;
\bibitem {H} Harrison, M : ``Introduction to Formal Language Theory'',\\ 
Addison-Wesley Ed., 1978;
\bibitem {HU} Hopcroft, J.E. - Ullman, J.D. : ``Introduction To Automata Theory'', Addison-Wesley Ed., 1979;
\end{thebibliography}
\end{document}